# Opto-mechanical inter-core cross-talk in multi-core fibers

HILEL HAGAI DIAMANDI, YOSEF LONDON[a], AND AVI ZADOK[*]

Faculty of Engineering and Institute of Nano-Technology and Advanced Materials, Bar-Ilan University, Ramat-Gan 5290002, Israel

## Abstract

Optical fibers containing multiple cores are widely regarded as the leading solution to the optical communications capacity crunch. The most prevalent paradigm for the design and employment of multi-core fibers relies on the suppression of direct coupling of optical power among cores. The cores, however, remain mechanically coupled. Inter-core, opto-mechanical cross-talk, among cores that are otherwise optically isolated from one another, is shown in this work for the first time. Light in one core stimulates guided acoustic modes of the entire fiber cladding. These modes, in turn, induce refractive index perturbations that extend across to other cores. Unlike corresponding processes in standard fiber, light waves in off-axis cores stimulate general torsional-radial guided acoustic modes of the cylindrical cross-section. Hundreds of such modes give rise to inter-core cross-phase modulation, with broad spectra that are quasi-continuous up to 1 GHz frequency. Inter-core cross-talk in a commercial, seven-core fiber is studied in both analysis and experiment. Opto-mechanical cross-talk is quantified in terms of an equivalent nonlinear coefficient, per acoustic mode or per frequency. The nonlinear coefficient may reach 1.9 [W×km]$^{-1}$, a value which is comparable with that of the intra-core Kerr effect in the same fiber.

---

[a] H. H. Diamandi and Y. London contributed equally to this work
[*] Corresponding author: Avinoam.Zadok@biu.ac.il

**Main Text**

**1. Introduction**

Multi-core fibers represent a major area of interest of the electro-optics community in recent years. They hold promise for supporting a large number of parallel optical communication channels, in space division multiplexing architectures [1-4]. The most common approach to the design and employment of these fibers is to try and reduce the residual coupling of optical power among cores as much as possible, in attempt to simplify system operation. Coupling may be suppressed by large physical separation between cores [5], heterogeneous cores with trenches of depressed cladding [6], or air holes [7]. The extent of residual coupling among cores has been studied in many works [5], which also addressed the effects of bending [8] and Kerr nonlinearity [9,10]. However, these works did not take into account opto-mechanical considerations. Although optically separated, the multiple cores are nevertheless assembled as part of a single, unified mechanical structure. The implications of this mechanical coupling are examined in this work. The results demonstrate that inter-core, opto-mechanical cross-phase modulation (XPM) may take place in multi-core fibers, even where direct optical coupling is very weak.

Opto-mechanical XPM is driven by guided acoustic modes of the entire fiber cladding [11]. Each mode is characterized by a cut-off frequency. Just above cut-off the mode is entirely transverse, and the axial component of its group velocity approaches zero [12-30]. The axial phase velocity, in contrast, approaches infinity. Hence for each mode there exists a specific frequency, very near cut-off, for which its axial acoustic phase velocity matches that of the guided optical modes of the fiber [12-30]. Therefore, the acoustic mode may be opto-mechanically stimulated by two co-propagating optical field components that are spectrally detuned by a proper frequency offset [12-30]. The phenomenon is known as guided acoustic waves Brillouin scattering (GAWBS) [12], forward stimulated Brillouin scattering [17], or Raman-like scattering by acoustic phonons [16].

GAWBS has been studied since 1985 [12]. The interactions were mapped in standard fibers [12,24], highly nonlinear fibers [17], solid-core photonic-crystal fibers [16,19,20] and micro-structured fibers [21-23], and their temperature and strain dependence was examined as well [25-29]. The effect was also observed in hollow-core fibers [30]. Recently, we have shown that GAWBS supports a new paradigm for the sensing of chemicals outside the cladding of an unmodified, standard optical fiber [24]. Measurements could be taken even though the guided light wave never came in contact with the substance under test [24].

The role of GAWBS in multi-core fibers may be understood as follows: the stimulated acoustic waves introduce perturbations to the local value of the refractive index, through the photo-elastic effect. These index variations extend across the entire cladding cross-section. When using standard single-mode fibers, the index perturbations may only be probed within the single core from which they are stimulated. In multi-core fibers, on the other hand, acoustically-induced index variations may affect light that is propagating in other cores as well. Opto-mechanical coupling among modes is extensively studied in various photonic devices [31,32], and was also addressed in few-order-mode, single-core fibers [33] and in dual nano-web, micro-structured fibers [21-23]. The phenomenon, however, was not yet investigated in multi-core fibers.

In the following we report a quantitative analytic and experimental study of inter-core, opto-mechanical coupling in a commercial, seven-core fiber. Our results show significant qualitative differences between GAWBS in a multi-core fiber, and corresponding processes in a standard, single-mode fiber. Guided acoustic waves scattering in standard single-mode fibers only involves radial modes, and one specific sub-category of torsional-radial modes, due to symmetry considerations [12,13]. The resulting spectra of probe wave modulation consist of discrete and sparse resonances, with comparatively large separations [12,17,24]. In contrast, light propagating in outer, off-axis cores of a multi-core fiber stimulates general, torsional-radial guided acoustic modes of the cylindrical cladding cross-section. Hundreds of such modes contribute to inter-core XPM, with spectra that are broad and quasi-continuous up to 1 GHz frequency. Broadband GAWBS spectra were previously observed in a hexagonal, single-core photonic crystal fiber [34].

Although considerably broader than GAWBS process in standard fiber, the bandwidth of the resulting inter-core XPM remains much narrower than the data rates of modern optical fiber communication. Hence the effect is unlikely to restrict the capacity of space-division multiplexing networks. On the other hand inter-core XPM may affect, and even serve for, other potential applications of multi-core fibers such as distributed chemical and shape sensors, opto-electronic oscillators, parametric amplifiers, or lasers operating at transverse super-modes across multiple cores. These prospects are addressed in the concluding discussion.

## 2. Analysis of Brillouin scattering by radial guided acoustic waves in multi-core fibers

The fiber used in this work consists of a central, inner core, and six outer cores that are equally spaced on a hexagonal grid. The centers of the outer cores are 35 µm away from the fiber center. The mode field diameter of the optical modes in all cores is specified as 6.4 ± 0.2 µm at 1550 nm wavelength. The optical power coupling between any pair of cores, in the fiber itself and in the fan-out units at its both ends, was verified as lower than -40 dB.

The mathematical analysis of inter-core XPM induced by GAWBS in multi-core fibers is provided in detail in the Supplementary Material. Only final expressions are given briefly below. We first address radial guided acoustic modes $R_{0m}$, where $m$ is an integer. These modes are stimulated by a pump wave that is propagating, for the time being, at the central, inner core. Inter-core XPM due to radial acoustic modes is characterized by a set of discrete resonances, in similarity to GAWBS in standard single-mode fiber. Radial modes are simpler to characterize quantitatively in experiment, hence their study also serves for the validation of our model. Treatment is extended to the more general case of pump waves in outer cores in section 4.

Let us denote the instantaneous power of the pump as a function of time $t$ by $P(t)$. The Fourier transform of $P(t)$ is $\tilde{P}(\Omega)$, where $\Omega$ is a radio-frequency (RF) variable. XPM is introduced to a second, optical probe wave. The Fourier components of opto-mechanical phase modulation due to mode $R_{0m}$ are given by [13]:

$$\delta\tilde{\phi}_{OM}^{(m)}(\Omega) = \frac{k_0}{4n^2 c \rho_0} \frac{Q_{ES}^{(m)} Q_{PE}^{(m)}}{\Gamma_m \Omega_m} L \frac{\tilde{P}(\Omega)}{j - 2(\Omega - \Omega_m)/\Gamma_m} \tag{1}$$

Here $\Omega_m$ is the cut-off frequency of mode $R_{0m}$, $\Gamma_m$ denotes the modal linewidth, $\rho_0$ and $n$ are the density and refractive index of silica respectively, $c$ is the speed of light in vacuum, $k_0$ is the vacuum wave-number of the probe wave, and $L$ is the length of the fiber. $Q_{ES}^{(m)}$ and $Q_{PE}^{(m)}$ are transverse overlap integrals that determine the electro-strictive stimulation of mode $R_{0m}$ and the photo-elastic modifications of the probe effective index by the same mode, respectively. Specific forms of Eq. (1) for probe waves in outer cores are derived in the Supplementary Material. The instantaneous phase modulation $\delta\phi_{OM}(t)$ may be obtained by the inverse Fourier transform of Eq. (1) and summation over $m$.

Based on Eq. (1), and following [35], we define an equivalent nonlinear coefficient for opto-mechanical XPM:

$$\gamma_{OM}^{(m)} \equiv \frac{k_0}{4n^2 c \rho_0} \frac{Q_{ES}^{(m)} Q_{PE}^{(m)}}{\Gamma_m \Omega_m} \qquad (2)$$

At the acoustic resonance frequencies we obtain the following relation between the power spectral density (PSD) of XPM and that of the pump power [35]:

$$\left| \delta \tilde{\phi}_{OM}^{(m)}(\Omega_m) \right|^2 = \left| \gamma_{OM}^{(m)} \right|^2 L^2 \left| \tilde{P}(\Omega_m) \right|^2 \qquad (3)$$

The units of $\gamma_{OM}^{(m)}$ are [W×km]$^{-1}$. The coefficient is mode-specific and core-specific, depends on the geometry of the fiber, and may also depend on the state of polarization (SOP) of the probe wave (see Supplementary Material). It is independent of the pump wave. The opto-mechanical coefficient holds an analogous role to that of the nonlinear coefficient $\gamma_{Kerr}$ associated with the intra-core Kerr effect, and the two might be compared.

**3. Experimental demonstration of inter-core cross-phase modulation by radial acoustic modes**

*3.1 Setup and measurement procedures*

A schematic illustration of the experimental setup is shown in Fig. 1 [16,24]. Light from a first laser diode source at 1553 nm wavelength was amplitude-modulated either by a continuous RF sine wave or by short and isolated periodic pulses, and used as a pump wave. The source linewidth was 100 kHz. Modulation of the pump serves for scanning the acoustic waves spectrum. The relative magnitude of XPM due to different acoustic modes is more easily measured with short pump pulses, since multiple modes may be addressed simultaneously. On the other hand, the absolute magnitudes of individual nonlinear coefficients $\gamma_{OM}^{(m)}$ are measured more precisely with sine-wave modulation.

Pulse modulation was carried out by a semiconductor optical amplifier (SOA) and an electro-optic modulator (EOM) connected in series [24]. The SOA provided a high modulation extinction ratio, which cannot be achieved in the EOM, but could not support short pulses. Pulses were first generated in the SOA with 5 ns duration and 1 µs period, and were then further modulated to 0.5 ns duration with the same period by the EOM. The pulse generators driving the SOA and EOM were synchronized. Pulses were amplified to peak power levels between 0.3-6.0 W by an erbium-doped fiber amplifier (EDFA). Sine-wave modulation of the pump was carried out using the EOM only. The sine-wave pump was amplified by the EDFA to average optical power levels between 20-60 mW. The pump wave was launched into the inner core of a 30 m-long, seven-core fiber under test, in one direction. A polarization scrambler along the input path of the pump wave was used to suppress the stimulation of polarization-sensitive, torsional-radial $TR_{2m}$ acoustic modes [11-13] (see also Supplementary Material).

The fiber under test was placed within a Sagnac loop [16,24] (Fig. 1). A probe wave from a second laser diode, at 1550 nm wavelength and of 10 kHz linewidth, was launched into either the inner core or an outer core of the same fiber, in both directions of the loop. A polarization controller (PC) was used to adjust the input SOP of the probe wave. Due to the wave-vector matching characteristics of GAWBS, opto-mechanical phase delay modifications $\delta\phi_{OM}(t)$ to the clockwise (CW) propagating probe wave accumulated over the entire fiber length [16,24]. In contrast, the counter-clockwise (CCW) propagating probe was subject to negligible nonlinear phase perturbation [16,24]. The acoustic waves therefore introduced non-reciprocal phase modulation of the probe wave. A second PC inside the loop was used to adjust the SOPs of the CW and CCW probe waves and the bias value of non-reciprocal phase delay $\phi_B$.

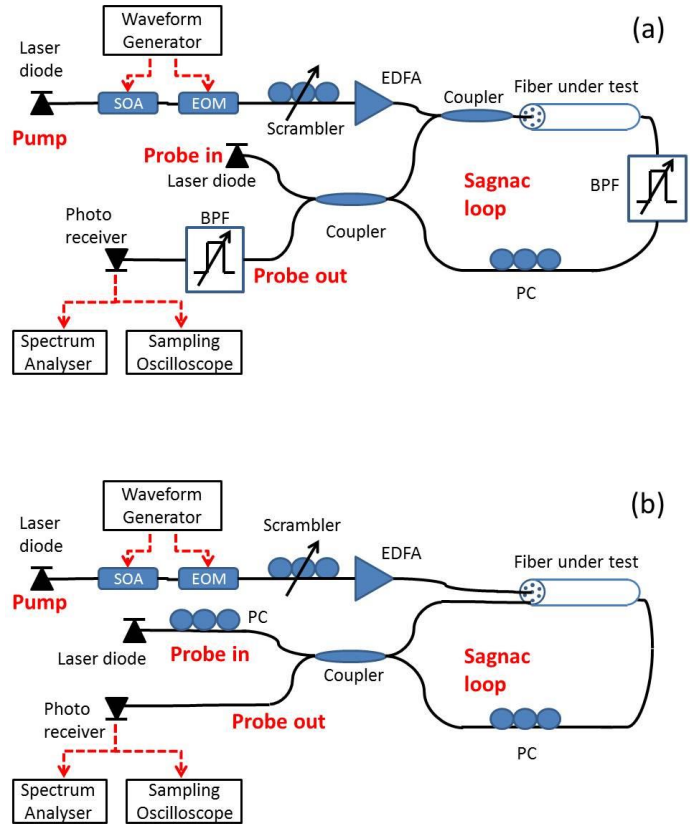

Fig. 1. Schematic illustrations of the experimental setup used to study opto-mechanical inter-core cross-phase modulation in a commercial seven-core fiber [16,24]. The pump wave propagated in the central core of the fiber under test, in one direction only. In some experiments the pump wave was modulated by pulses of 0.5 ns duration and 1 µs period, using a semiconductor optical amplifier (SOA) and an electro-optic modulator (EOM) connected in series. Pulses were amplified to peak power levels between 0.3-6.0 W by an erbium-doped fiber amplifier (EDFA). In other experiments, the pump wave was modulated by continuous RF sine waves using the EOM only, and amplified by the EDFA to average power levels of 20-60 mW. A continuous probe wave propagated in either the inner core (panel (a)) or an outer core (panel (b)) of the same fiber, in both directions, in a Sagnac loop configuration. When the probe propagated in the inner core alongside the intense pump (panel (a)), optical bandpass filters (BPFs) were used to block the pump wave from reaching the loop output. GAWBS driven by the pump pulses induced non-reciprocal phase delay perturbation to the probe wave.

The non-reciprocal phase delay was converted to an intensity signal upon detection of the probe wave at the loop output. When the probe wave propagated at the inner core alongside the intense pump wave, optical bandpass filters were used to block the pump from reaching the detector (see Fig. 1(a)). The bandpass filters were not needed when the probe wave propagated in an outer core. The detector output voltage $\delta V(t)$ was either observed with an RF spectrum analyzer, or sampled by a real-time digitizing oscilloscope of 6 GHz bandwidth. Traces recorded by the oscilloscope were averaged over 1,024 to 4,096 repetitions and analyzed using offline signal processing.

We may relate the instantaneous detector voltage to the opto-mechanical XPM of the probe wave as follows:

$$\delta V(t) \approx \tfrac{1}{2} V_{\max} \delta\phi_{OM}(t) \tag{4}$$

Here $V_{\max}$ denotes the maximum output voltage of the detector, which is obtained when the CW and CCW probe waves interfere constructively. Eq. (4) is valid when the following conditions are met: the SOPs of CW and CCW probe waves at the detector input are parallel, $\phi_B = \pi/2$ and $\delta\phi_{OM}(t) \ll \pi$. However, the measured $\delta V(t)$ could be smaller than that of Eq. (4) if the two SOPs are not aligned and/or if $\phi_B \neq \pi/2$. The SOPs of the probe waves vary in an uncontrollable manner along the fiber under test. In addition, the magnitude of $\delta\phi_{OM}(t)$ in the outer cores is polarization-dependent (see Supplementary Material). The two PCs in the experimental setup were adjusted in attempt to maximize $\delta V(t)$. The alignments have little influence on relative measurements of the normalized XPM spectra $\left|\delta\tilde{\phi}_{OM}(\Omega)\right|^2$, but may affect quantitative, absolute measurements of $\gamma_{OM}^{(m)}$.

*3.2 Results*

Measurements of probe waves at the inner core were used for calibration and validation purposes. Figure 2(a) shows a measurement of $\delta V(t)$ obtained using short pump pulses. The trace consists of a series of impulses, separated by the acoustic propagation delay from the inner core to the outer boundary of the cladding and back: $t_0 = 20.8$ ns [24]. The sampled trace was digitally filtered to select individual modal contributions $\delta V^{(m)}(t)$ (see an example in Fig. 2(b)). The passbands of the digital filters were centered at $\Omega_m/(2\pi)$, and their full-widths at half-maximum were 45 MHz. Modal linewidths $\Gamma_m$, used in calculations of Eq. (1), were extracted from the exponential decay rates of filtered traces. Figure 2(c) shows the normalized PSD of the unfiltered output voltage, $\left|\delta\tilde{V}(\Omega)\right|^2$, alongside the calculated normalized probe wave XPM spectrum $\left|\delta\tilde{\phi}_{OM}(\Omega)\right|^2$. The measurement results are in very good agreement, both with the analysis and with previous measurements of GAWBS in single-mode fibers [12,17,24]. The cladding diameter was fitted based on the measured resonances: $2a = 125.6$ µm [12,13,24].

Figure 3(a) shows the measured $\delta V(t)$ with the probe wave moved to an outer core, obtained using short pulses. Opto-mechanical XPM due to pump pulses in the inner core is evident. The temporal trace retains its fundamental periodicity of $t_0$. However, the separation between adjacent impulses is no longer fixed, since the distance from the outer cores to the cladding boundary is not exactly equal to the distance from the outer cores to the fiber axis. A filtered trace of $\delta V^{(8)}(t)$ is shown in Fig. 3(b). Figure 3(c) shows measurements of the peak-to-peak magnitude of $\delta V(t)$, as a function of the average pump power. The pump wave in this experiment was modulated by a sine-wave of frequency $\Omega_8 = 2\pi \cdot 369.2$ MHz. A linear proportion is observed, as anticipated.

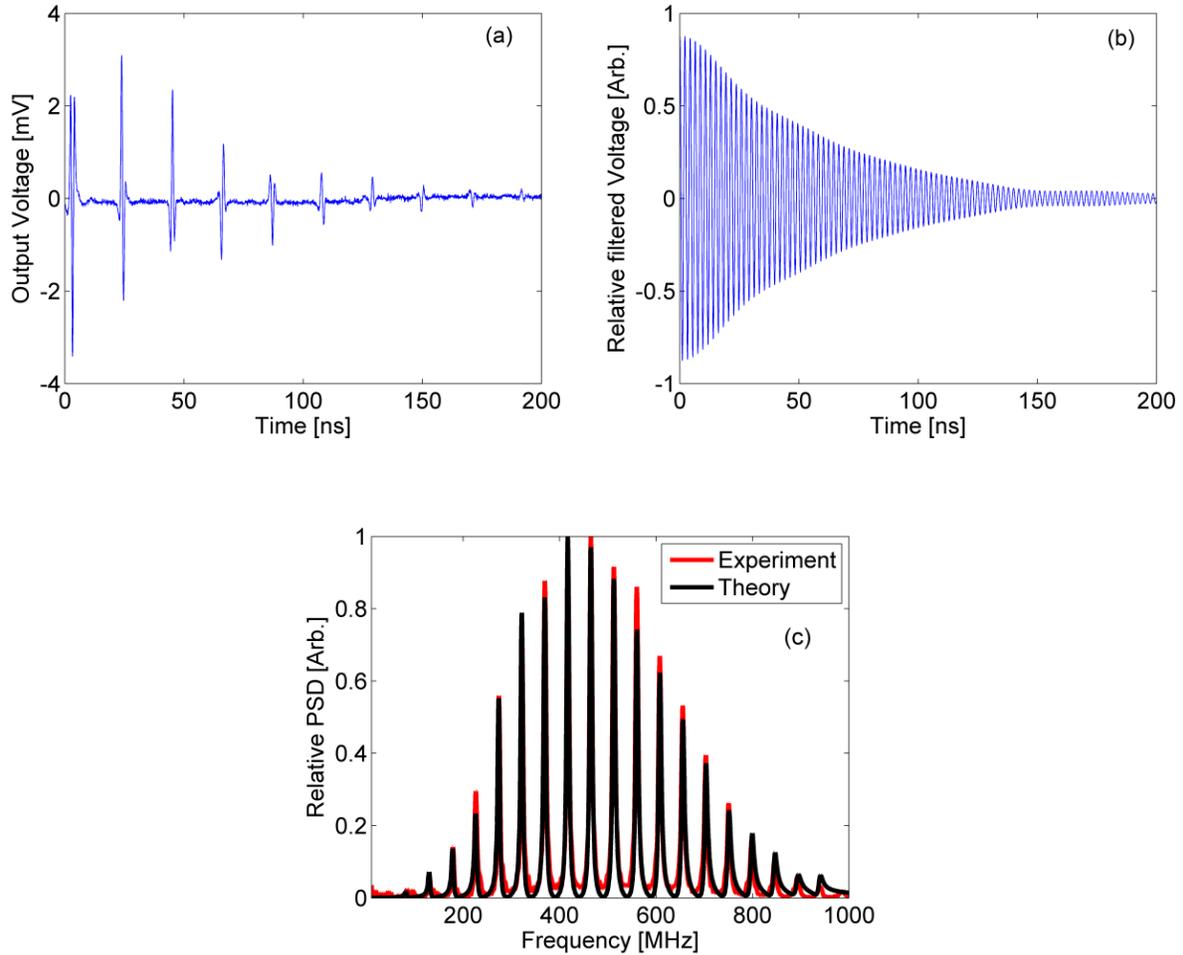

Fig. 2. (a) – Output voltage of the detected probe wave as a function of time, following propagation in the inner core. Pump pulses were 0.5 ns-long, with peak power of 3.5 W. (b) – The trace of panel (a), digitally filtered to select the contribution of acoustic mode $R_{0,10}$ at resonance frequency $\Omega_{10} = 2\pi \cdot 464$ MHz. The modal oscillations decay exponentially with a linewidth $\Gamma_{10} = 2\pi \cdot 6.4$ MHz. (c) – Measured (red) and calculated (black) normalized PSDs of the probe wave opto-mechanical phase modulation. Good agreement between measurements, analysis and previous studies is achieved.

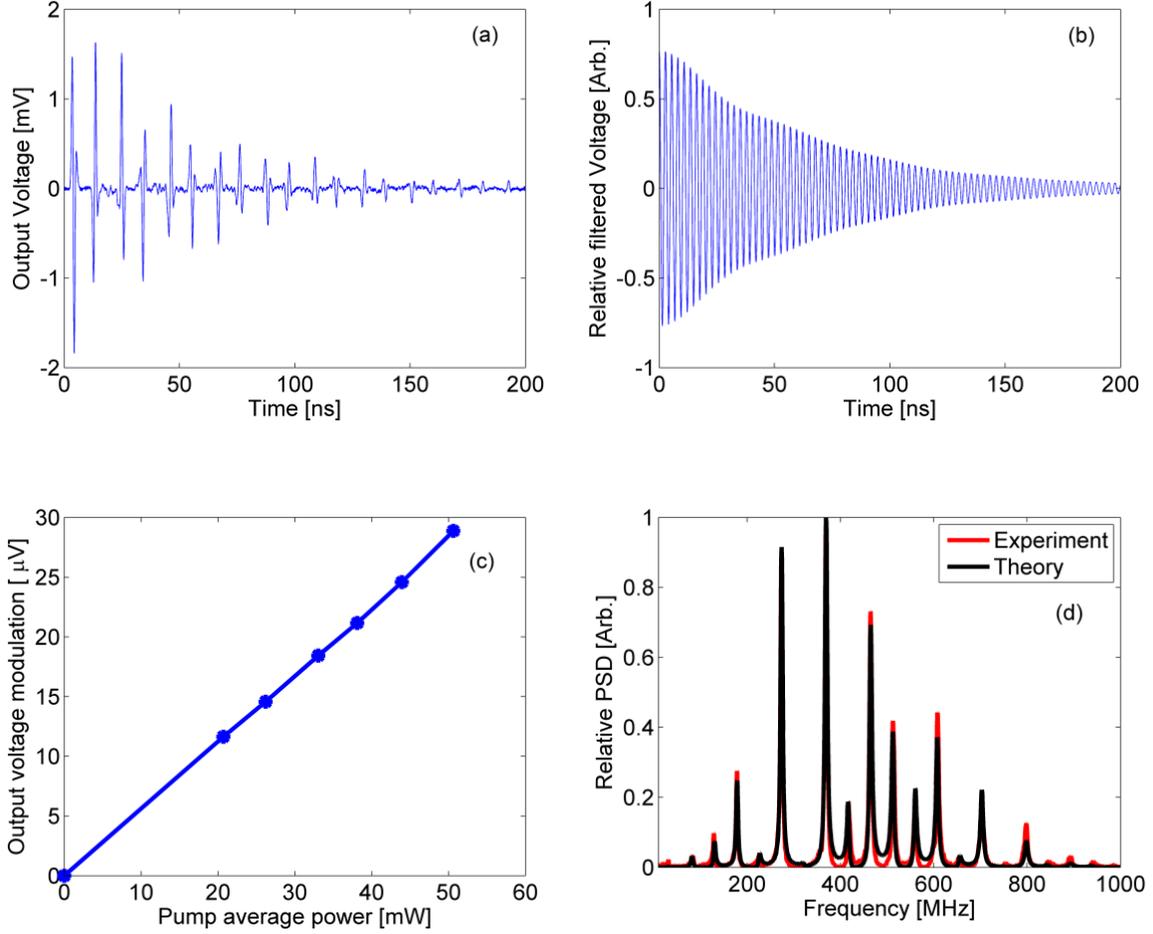

Fig. 3. (a) – Output voltage of the detected probe wave as a function of time, following propagation in an outer core. Pump pulses were 0.5 ns-long, with peak power of 3.5 W. (b) – The trace of panel (a), digitally filtered to select the contribution of mode $R_{08}$ at resonance frequency $\Omega_8 = 2\pi \cdot 369.2$ MHz. (c) – Measured peak-to-peak magnitude of the output voltage, as a function the average pump power. The pump was modulated by a sine-wave of frequency $\Omega_8$. A linear relation is observed as expected. (d) – Measured (red) and calculated (black) normalized PSDs of the probe wave phase modulation. The irregular spectrum observed is fully accounted for by the analysis

The normalized PSD of the unfiltered output probe wave of Fig. 3(a), $\left|\delta \tilde{V}(\Omega)\right|^2$, is shown in Fig. 3(d). In contrast to Fig. 2(c), the output probe spectrum following propagation in an outer core is highly irregular: large differences are observed among the PSDs in adjacent resonances. The numerically-calculated, normalized PSD of inter-core XPM $\left|\delta \tilde{\phi}_{OM}(\Omega)\right|^2$ is shown in Fig. 3(d) as well. Excellent agreement is obtained between measured and calculated normalized spectra. The

differences among modes are due to the photo-elastic overlap integrals $Q_{PE}^{(m)}$, as illustrated in Fig. 4(a) and 4(b). The transverse overlap between the profile of GAWBS-induced dielectric constant variations and the probe optical mode across outer cores is substantial for some values of $m$, but almost vanishes for others. The PSD of the detected probe wave provides clear indication for opto-mechanical XPM.

The largest inter-core XPM was observed for mode $R_{08}$. The expected magnitude of the opto-mechanical nonlinear coefficient for that mode was numerically calculated using Eq. (2). Values of $\left|\gamma_{OM}^{(8)}\right| = 1.9 \pm 0.1$ [W×km]$^{-1}$ and $0.9 \pm 0.1$ [W×km]$^{-1}$ are predicted for two orthogonal SOPs of the probe wave, respectively. Uncertainty is due to tolerance in the specifications of the optical mode field diameter. A corresponding experimental estimate could be obtained using the measured XPM spectrum, Eq. (3) and Eq. (4):

$$\left|\left\langle \gamma_{OM}^{(8)} \right\rangle_z\right| = 2\left|\delta\tilde{V}(\Omega_8)\right|/\left[V_{max} L \left|\tilde{P}(\Omega_8)\right|\right] \qquad (5)$$

The nonlinear coefficient in Eq. (5) is estimated based on the accumulation of XPM over the entire length of the fiber. Since the SOP of the probe wave varies arbitrarily along the fiber, the result represents a position-averaged value, noted above by $\langle \ \rangle_z$. As suggested earlier, measurements at the output of the Sagnac loop varied significantly with the adjustment of PCs. The largest value obtained was $\left|\left\langle \gamma_{OM}^{(8)} \right\rangle_z\right| = 1.3 \pm 0.2$ [W×km]$^{-1}$. This measurement is between the two calculated values of $\left|\gamma_{OM}^{(8)}\right|$ noted above, as could be expected. Experimental error is primarily due to uncertainty in the pump power that was actually coupled into the multi-core fiber. The end-to-end coupling losses from standard fiber to the multi-core fiber and back were measured as 2.5 dB, however the division of these losses between the two interfaces is unknown. Hence the pump power could not be determined with better precision.

Comprehensive numerical analysis of the probe wave propagation in the Sagnac loop, subject to opto-mechanical XPM and many realizations of arbitrary fiber birefringence [36], is reported in the Supplementary Material. The analysis replicates the experimental procedure. The calculated value of $\left|\left\langle \gamma_{OM}^{(8)} \right\rangle_z\right|$ was 1.4 ± 0.2 [W×km]$^{-1}$. The uncertainty represents the standard deviation among the results obtained for hundreds of different fiber realizations. The measured nonlinear coefficient is therefore in good agreement with analysis. The observed values are on the same order of magnitude as $\gamma_{Kerr}$ in each core, which is estimated as 4 [W×km]$^{-1}$ based on the fiber mode field diameter.

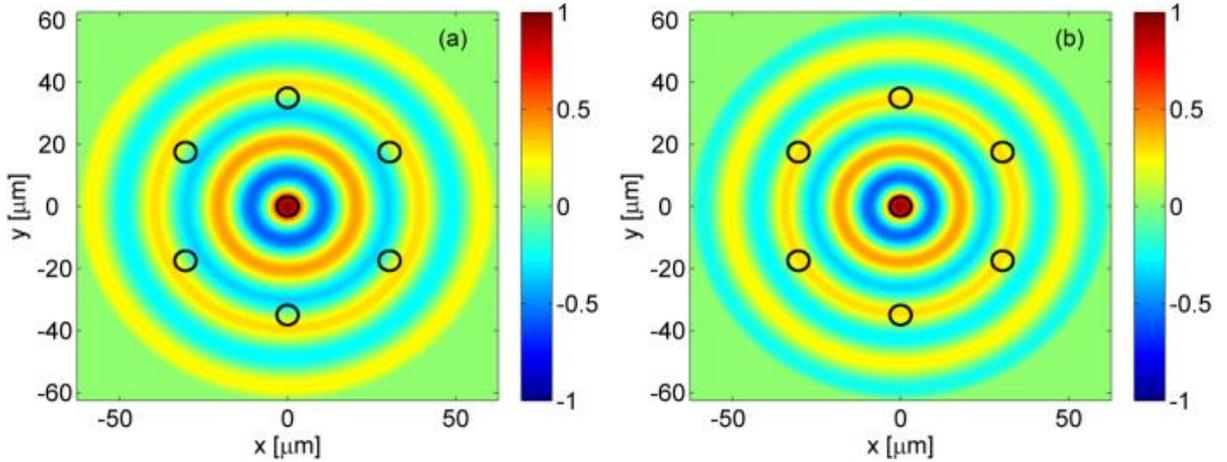

Fig 4. Simulated, normalized transverse profiles of the photo-elastic index variations as a function of Cartesian coordinates $x$ and $y$. The index changes were calculated for light that is polarized along the local azimuthal direction (see Supplementary Material). The cores of the fiber are noted in black circles. Panels (a) and (b) correspond to modes $R_{07}$ and $R_{08}$, respectively. The index perturbation for $R_{07}$ ($\Omega_7 = 2\pi\cdot 322$ MHz) changes sign within the outer cores, leading to effective cancellation of the photo-elastic overlap integral $Q_{PE}^{(7)}$. The corresponding profile for mode $R_{08}$ ($\Omega_8 = 2\pi\cdot 369.2$ MHz) passes through a maximum within outer cores

## 4. Brillouin scattering by general torsional-radial guided acoustic waves

Light in a central, axis-symmetric core of a fiber with a cylindrical cross-section may only stimulate two categories of guided acoustic modes: the radial modes $R_{0m}$ addressed thus far, and torsional-radial modes with two-fold azimuthal symmetry, noted as $TR_{2m}$ [11-13]. The stimulation of the latter category was deliberately suppressed in the experiments of section 3, by scrambling the polarization of the pump wave [11-13]. The fiber cross-section supports a broad variety of general torsional-radial guided acoustic modes, denoted as $TR_{pm}$, where $p \geq 0$ is any integer [37-39]. These modes cannot be addressed by GAWBS processes in standard, single-mode fibers. However, they may be stimulated by pump light which propagates in an outer, off-axis core of a multi-core fiber.

An extended analysis of GAWBS in multi-core fibers, which accounts for pump waves in outer cores, is reported in detail in the Supplementary Material. The analysis suggests that several hundreds of guided torsional-radial acoustic modes contribute to GAWBS in the seven-core fiber under test. Figure 5 shows an example of the normalized transverse profile of the GAWBS-induced perturbation $\delta\tilde{\varepsilon}$ to the local dielectric tensor of the fiber, due to mode $TR_{18,14}$. The acoustic cut-off frequency of the particular mode is $\Omega_{18,14} = 2\pi \cdot 526$ MHz. Local maxima of the dielectric tensor perturbation are in spatial overlap with the outer cores, suggesting effective inter-core XPM through that mode.

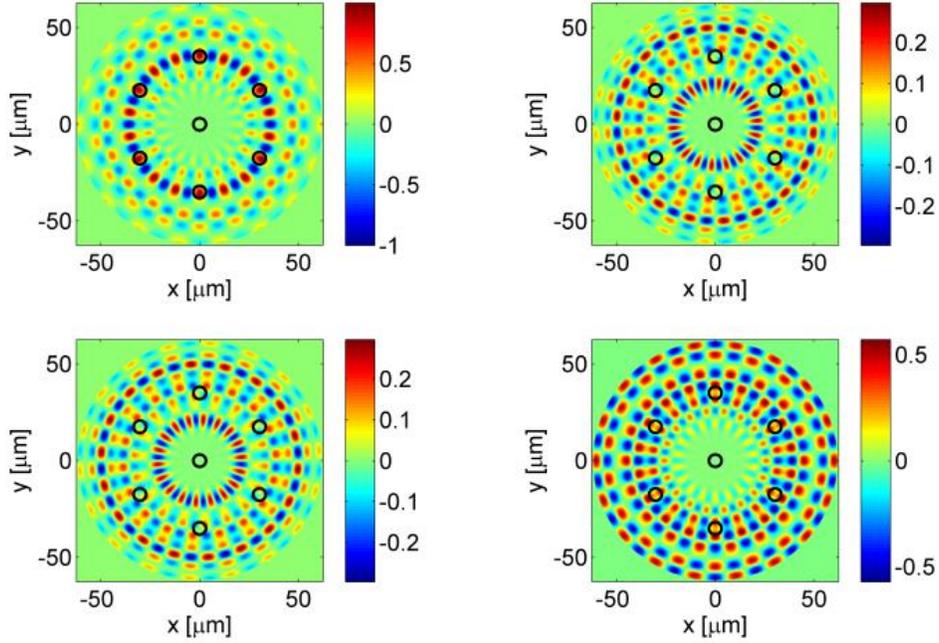

Fig 5. Example of the calculated normalized transverse profile of the GAWBS-induced dielectric tensor perturbation $\delta\tilde{\boldsymbol{\varepsilon}}$. The profile correspond to the torsional-radial guided acoustic mode $TR_{18,14}$. The cut-off frequency of the particular mode is $\Omega_{18,14} = 2\pi \cdot 526$ MHz. The dielectric tensor perturbations are presented in polar coordinates $r, \varphi$. Top left: $\delta\tilde{\varepsilon}_{rr}$. Top right: $\delta\tilde{\varepsilon}_{r\varphi}$. Bottom left: $\delta\tilde{\varepsilon}_{\varphi r}$. Bottom right: $\delta\tilde{\varepsilon}_{\varphi\varphi}$.

The resonant spectra of many torsional-radial modes are in significant overlap. Due to that overlap, the definition of mode-specific nonlinear coefficients for the inter-core GAWBS process (as in Eq. (2)) is less convenient for $TR_{pm}$ modes. Instead, we may define two frequency-dependent opto-mechanical nonlinear coefficients, which take into account the combined effects of all guided acoustic modes:

$$\tilde{\gamma}_{OM}^{(i)}(\Omega) \equiv \frac{k_0}{4n^2 c\rho_0} \tilde{H}_{OM}^{(i)}(\Omega). \tag{6}$$

Here $\tilde{H}_{OM}^{(i)}(\Omega)$, $i = 1, 2$ are the eigen-values of a frequency-dependent opto-mechanical tensor, which is defined and analyzed in detail in the Supplementary Material. The coefficients $\tilde{\gamma}_{OM}^{(i)}(\Omega)$ quantify the XPM of probe waves that are polarized along the two respective eigen-vectors of the same tensor:

$$\delta\tilde{\phi}^{(i)}(\Omega) = \gamma_{OM}^{(i)}(\Omega)\tilde{P}(\Omega)L. \tag{7}$$

Figure 6 shows an example of the calculated position-averaged coefficient $\left|\langle\tilde{\gamma}_{OM}(\Omega)\rangle_z\right| \equiv \frac{1}{2}\left|\tilde{\gamma}_{OM}^{(1)}(\Omega) + \tilde{\gamma}_{OM}^{(2)}(\Omega)\right|$. The simulation was carried out for a pump wave in one outer core, and a probe wave in an adjacent outer core. Modes of azimuthal orders $0 \leq p \leq 36$ were considered. The analysis suggests a scattering spectrum that is broad and quasi-continuous, up to 1 GHz frequency. Nonlinear coefficients as high as 1.8 [W×km]$^{-1}$ are calculated at certain frequencies. The nonlinear coefficient is larger than 0.5 [W×km]$^{-1}$ for all frequencies between 200 MHz and 800 MHz. Similar spectra were calculated for probe waves at other outer cores. XPM in the opposite outer core remains significant even up to 1.5 GHz.

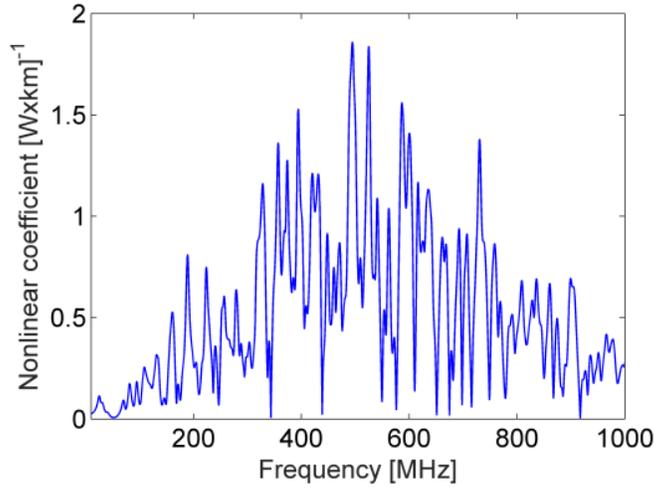

Fig. 6. Calculated equivalent nonlinear coefficient of guided acoustic waves Brillouin scattering between two adjacent outer cores in a seven-core fiber, as a function of radio-frequency. The scattering spectrum consists of contributions of hundreds of torsional-radial acoustic modes. Azimuthal orders $p$ between 0 and 36 were considered

Measurements of GAWBS were carried out using the setup of Fig. 1(b), with the short pump pulses moved to an outer core and the probe wave propagated in an adjacent outer core. Figure 7(a) shows an example of the instantaneous detector output $\delta V(t)$. Unlike Fig. 2(a) and Fig. 3(a), the temporal trace is noise-like and irregular. Figure 7(b) shows the measured normalized PSD of the output probe wave $|\delta \tilde{V}(\Omega)|^2$. Inter-core, opto-mechanical XPM is observed up to 750 MHz frequency, limited by the bandwidth of the pump pulses. One of the strongest spectral peaks matches $\Omega_{18,14}$. Similar spectra were obtained when the probe wave was moved to other outer cores. The qualitative prediction for broad inter-core cross-talk is therefore corroborated by experiment. Measured spectra are markedly different from those obtained through $R_{0m}$ modes only (see Fig. 2(c) and Fig. 3(d)), which consist of discrete, sparse resonances with comparatively large separations. The results provide a first demonstration of GAWBS involving general $TR_{pm}$ modes in fibers with cylindrically-symmetric mechanical structure. Broadband GAWBS spectra due to a large number of modes were previously demonstrated in a single-core, photonic crystal fiber of hexagonal structure [34].

Unlike the case of sparse radial modes addressed earlier, the details of the calculated and measured XPM spectra due to hundreds of overlapping $TR_{pm}$ modes do not fully agree. A possible source for discrepancy is depolarization. The probe wave modulation is polarization-dependent, with principal axes that vary with frequency. Hence the visibility of interference between CW and CCW probe waves may become frequency-dependent, and distort the measured spectra. In addition, torsional-radial modes of azimuthal orders $p > 36$ may also contribute to inter-core cross-talk. Last, the doping profiles of the cores may modify their mechanical properties, and thereby affect the exact resonance frequencies and transverse profiles of high-order torsional-radial modes. Acoustic guiding in the core is known, for example, in backwards stimulated Brillouin scattering. This effect was not included in the analysis.

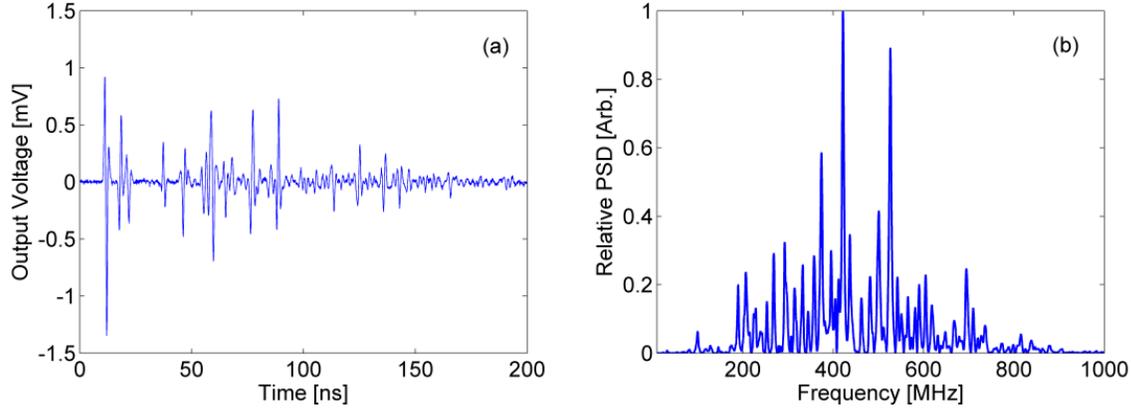

Fig. 7. (a) - Output voltage of the detected probe wave as a function of time, at the output of the Sagnac loop, following propagation in an outer core. Short pump pulses of 10 W peak power and 0.7 ns duration propagated in an adjacent outer core. (b) Measured nonlinear power spectral density of the output probe wave. A broad, quasi-continuous phase modulation spectrum is observed up to a frequency of 750 MHz, limited by the bandwidth of pump pulses. The spectrum is in marked contrast to those of radial modes only (see Fig. 3(d)).

## 5. Discussion

In this work, we have studied inter-core GAWBS processes in a commercial, seven-core fiber. The results provide a first demonstration of cross-talk among cores that are otherwise optically isolated from one another. This cross-talk mechanism was not considered before. The specific case of a polarization-scrambled pump wave at the central core was examined first. Cross-talk in this case takes place through radial guided acoustic modes. Analysis showed that opto-mechanical XPM of probe waves propagating in all cores may be expected. The modulation spectra consist of a series of narrowband resonances. The magnitude of the effect is quantified in terms of an equivalent nonlinear coefficient per each acoustic mode [35]. Calculations suggest that the nonlinear coefficient can be on the same order of magnitude as that of the intra-core, Kerr nonlinearity. Inter-core XPM subject to these conditions was observed experimentally. Excellent quantitative agreement was obtained between model and measurement, in both magnitude and PSD of the inter-core cross-talk.

The study was then extended to the more general case of pump waves that propagate in outer, off-axis cores. Due to the removal of radial symmetry, GAWBS in this case becomes a fundamentally different process. Hundreds of general torsional-radial acoustic modes contribute to inter-core cross-talk, with a broad spectrum that reaches 1 GHz. Broad modulation spectra were also observed experimentally. Although general torsional-radial guided acoustic waves in cylindrical fiber are long known [37-39], they cannot be observed through Brillouin scattering in standard fiber. The results demonstrate that multi-core fibers open new possibilities for the study of fiber opto-mechanics.

The results carry two main messages: a) One may not simply assume that optical isolation alone would guarantee the cross-talk-free operation of multi-core fibers in all conditions; and b) Guided acoustic waves Brillouin scattering in multi-core fibers is not as narrowband as might have been anticipated based on corresponding processes in single-mode fibers. That being said, inter-core opto-mechanical XPM is unlikely to restrict the capacity of currently-proposed space-division-multiplexing optical fiber communication networks. The bandwidth of inter-core GAWBS remains much narrower than communication data rates. Hence only a fraction of the spectral components of broadband data would contribute towards opto-mechanical nonlinearities, while the entire power manifests in the Kerr effect. Even though opto-mechanical XPM would scale with the number of cores, we expect that Kerr nonlinearity would still impose the more stringent limitation on the performance of most space-division multiplexing communication systems.

On the other hand, our analysis suggests that opto-mechanical cross-talk due to a single pair of cores is of comparable magnitude to the intra-core Kerr effect for frequencies up to the order of 1

GHz. GAWBS might therefore become the dominant nonlinearity when the transmission bandwidth is on that order. This could be the case, for example, in sub-carrier-multiplexed radio-over-fiber transmission [40] or in microwave-photonic applications of multi-core fibers, as proposed in [41]. While multi-core fibers are currently pursued primarily for the purpose of high-rate communications, their use may diversify as they become more widely available.

One application of highly coherent light in multi-core fibers is in fiber lasers [42-45]. Multi-core fiber lasers exhibit self-organization of spatial super-modes across several cores. The formation and stability of these modes depend on the extent of coupling among cores [42-45]. Coupling might be modified by inter-core GAWBS processes. Opto-mechanical crosstalk might impede, or perhaps assist, the propagation of transverse super-modes in high-power fiber lasers. GAWBS is also identified as a source of noise in parametric fiber amplifiers [46]. Parametric gain is being considered for phase-sensitive amplification in coherent optical communication [47]. In some cases, parametric amplifiers employ carefully designed doping profiles to suppress the guiding of acoustic modes in the core, and elevate the threshold of backwards stimulated Brillouin scattering at 10-11 GHz [47]. However, this solution path is inapplicable to GAWBS. Parametric amplification in space-division-multiplexing networks over multi-core fibers may have to take inter-core GAWBS into consideration.

While inter-core GAWBS can be detrimental from the standpoints of certain applications, the same effect could become very useful in other contexts. One such example is the recently-proposed protocol for chemical sensing outside the cladding [24]. In that approach, changes in the modal linewidth of guided acoustic waves due to the acoustic impedance at the cladding boundary are monitored. The method works around a long-time difficulty of fiber sensors: the lack of spatial overlap between light that is confined to the core and substances outside the cladding of unmodified fibers. Thus far, the method has been demonstrated over standard single-mode fiber [24], however its extension to multi-core fibers would bring about significant added values. First, spatial separation between pump and probe would improve the measurement signal-to-noise ratio and precision. Second, use of fibers consisted of heterogeneous, dissimilar cores would restrict the pump-probe interactions to the walk-off distances among different optical modes, and may give rise to distributed analysis. This prospect is highly sought-after by the optical fiber sensors community. Last but not least, simultaneous opto-mechanical measurements in multiple cores can also support advanced shape sensing [48,49]. Shape-sensing in multi-core fibers relies on narrowband optical signals.

Another potential application is in opto-electronic oscillators [50]. In such oscillators light is amplitude-modulated by an RF waveform, and propagates along a section of fiber. The waveform is detected at the output of the fiber, and the recovered RF signal is fed back to modulate the optical input. When the feedback gain is sufficiently high, stable self-sustained RF oscillations may be achieved [50]. Opto-electronic oscillators provide RF tones with extremely low phase noise, and they are pursued for applications in coherent communication, radars and precision metrology. We have recently demonstrated an opto-electronic oscillator which makes use of GAWBS in standard single-mode fiber as the sole mechanism for frequency selectivity [51]. Here too, the employment of multi-core fibers would reduce phase noise due to separation between pump and probe, and provide larger freedom in the choice of radio-frequencies.

In conclusion, multi-core optical fibers provide a rich playground for the study of opto-mechanics. The applications and implications of the principles demonstrated in this work are the subjects of ongoing research.

**Funding**

The authors acknowledge the financial support of the European Research Council (ERC), grant no. H2020-ERC-2015-STG 679228 (L-SID); and of the Israeli Science Foundation (ISF) under grant number 1665-14.

**Akcnowledgements**

The authors thank Dr. Yair Antman of Bar-Ilan University for his assistance in the analysis and experiments.

See Supplement 1 for supporting content.

**References**

1. E. D. Ellis, J. Zhao, and D. Cotter, "Approaching the non-linear Shannon limit," J. Lightwave Technol. **28**, 423-433 (2010).

2. R. J. Essiambre, G. J. Foschini, G. Kramer, and P. J. Winzer, "Capacity limits of information transport in fiber-optic networks," Phys. Rev. Lett. **101**, 163901 (2008).

3. D. J. Richardson, J. M. Fini, and L. E. Nelson, "Space-division multiplexing in optical fibres," Nat. Photon. **7**, 354-362 (2013).

4. R. Ryf, S. Randel, A. H. Gnauck, C. Bolle, A. Sierra, S. Mumtaz, M. Esmaeelpour, E. C. Burrows, R. J. Essiambre, P. J. Wizner, D. W. Peckham, A. H. McCurdy, and R. Lingle, "Mode-division multiplexing over 96 km of few-mode fiber using coherent 6 × 6 MIMO processing," J. Lightwave Technol. **30**, 521-531 (2012).

5. M. Koshiba, K. Saitoh, K. Takenaga, and S. Matsuo, "Multi-core fiber design and analysis: coupled-mode theory and coupled-power theory," Opt. Express **19**, B102-B111 (2011).

6. J. Tu, K. Saitoh, M. Koshiba, K. Takenaga, and S. Matsuo, "Design and analysis of large-effective-area heterogeneous trench-assisted multi-core fiber," Opt. Express **20**, 15157-15170 (2012).

7. T. Watanabe and Y. Kokubun, "Ultra-large number of transmission channels in space division multiplexing using few-mode multi-core fiber with optimized air-hole-assisted double-cladding structure," Opt. Express **22**, 8309-8319 (2014).

8. T. Hayashi, T. Sasaki, E. Sasaoka, K. Saitoh, and M. Koshiba, "Physical interpretation of intercore crosstalk in multicore fiber: effects of macrobend, structure fluctuation, and microbend," Opt. Express **21**, 5401-5412 (2013).

9. A. Macho, M. Morant, and R. Llorente, "Experimental evaluation of nonlinear crosstalk in multi-core fiber," Opt. Express **23**, 18712-18720 (2015).

10. C. Antonelli, M. Shtaif, A. and Mecozzi, "Modeling of Nonlinear Propagation in Space-Division Multiplexed Fiber-Optic Transmission," J. Lightwave Technol. **34**, 36-54 (2016).

11. S. W. Rienstra and A. Hirschberg, An Introduction to Acoustics (Eindhoven University of Technology, the Netherlands, 2015), Chapter 7.

12. R. M. Shelby, M. D. Levenson, and P. W. Bayer, "Guided acoustic-wave Brillouin scattering," Phys. Rev. B **31**, 5244-5252 (1985).

13. A. S. Biryukov, M. E. Sukharev, and E. M. Dianov, "Excitation of sound waves upon propagation of laser pulses in optical fibers," IEEE J. Quant. Elect. **32**, 765-775 (2002).

14. P. St. J. Russell, D. Culverhouse, and F. Farahi, "Experimental observation of forward stimulated Brillouin scattering in dual-mode single-core fibre," Elect. Lett. **26**, 1195-1196 (1990).

15. P. St. J. Russell, D. Culverhouse, and F. Farahi, "Theory of forward stimulated Brillouin scattering in dual-mode single-core fibers," IEEE J. Quant. Elect. **27**, 836-842 (1991).

16. M. S. Kang, A. Nazarkin, A. Brenn, and P. St. J. Russell, "Tightly trapped acoustic phonons in photonic crystal fibers as highly nonlinear artificial Raman oscillators," Nat. Physics **5**, 276-280 (2009).

17. J. Wang, Y. Zhu, R. Zhang, and D. J. Gauthier, "FSBS resonances observed in standard highly nonlinear fiber," Opt. Express **19**, 5339-5349 (2011).

18. E. Peral and A. Yariv, "Degradation of modulation and noise characteristics of semiconductor lasers after propagation in optical fiber due to a phase shift induced by stimulated Brillouin scattering," IEEE J. Quant. Elect. **35**, 1185–1195 (1999).

19. P. Dainese, P. St. J. Russell, N. Joly, J. C. Knight, G. S. Wiederhecker, H. L. Fragnito, V. Laude, and A. Khelif, "Stimulated Brillouin scattering from multi-GHz-guided acoustic phonons in nanostructured photonic crystal fibres," Nat. Physics **2**, 388-392 (2006).

20. M. Pang, W. He, X. Jiang, P. St. J. Russell, "All-optical bit storage in a fibre-laser by optomechanically bound states of solitons," Nat. Photon. **10**, 454 (2016).

21. J. R. Koehler, A. Butsch, T. G. Euser, R. E. Noskov, P. St. J. Russell, "Effects of squeezed film damping on the optomechanical nonlinearity in dual-nanoweb fiber," Appl. Phys. Lett. **103**, 221107 (2013).

22. A. Butsch, J. R. Koehler, R. E. Noskov, P. St. J. Russell, "CW-pumped single-pass frequency comb generation by resonant optomechanical nonlinearity in dual-nanpweb fiber," Optica **1**, 158 (2014).

23. J. R. Koehler, R. E. Noskov, A. A. Sukhorukov, A. Butsch, D. Novoa, P. St. J. Russell, "Resolving the mystery of milliwatt-threshold opto-mechanical self-oscillations in dual-nanoweb fiber," Appl. Phys. Lett. Photon. **1**, 056101 (2016).

24. Y. Antman, A. Clain, Y. London and A. Zadok, "Optomechanical sensing of liquids outside standard fibers using forward stimulated Brillouin scattering," Optica **3**, 510-516 (2016).

25. Y. Tanaka and K. Ogusu "Temperature coefficient of sideband frequencies produced by depolarized guided acoustic-wave Brillouin scattering," IEEE Photon. Technol. Lett. **10**, 1769–1771 (1998).

26. T. Matsui, K. Nakajima, T. Sakamoto, K. Shiraki, and I. Sankawa, "Structural dependence of guided acoustic-wave Brillouin scattering spectral in hole-assisted fiber and its temperature dependence," Appl. Opt. **46**, 6912–6917 (2007).

27. E. Carry, J. C. Beugnot, B. Stiller, M. W. Lee, H. Maillotte, and T. Sylvestre, "Temperature coefficient of the high-frequency guided acoustic mode in a photonic crystal fiber," Appl. Opt. **50**, 6543–6547 (2011).

28. Y. Tanaka and K. Ogusu, "Tensile-strain coefficient of resonance frequency of depolarized guided acoustic-wave Brillouin scattering," IEEE Photon. Technol. Lett. **11**, 865–867 (1999).

29. Y. Antman, Y. London and A. Zadok, "Scanning-free characterization of temperature dependence of forward stimulated Brillouin scattering resonances," in Proceedings of Optical Fiber Sensors 25 Conference (OFS-25), Proc. SPIE **9634**, 96345C (2015).

30. W. H. Renninger, R. O. Behunin, and P. T. Rakich, "Guided-wave Brillouin scattering in air," Optica **3**, 1316-1319 (2016).

31. M. Tomes and T. Carmon, "Photonic micro-electromechanical systems vibrating at X-band (11-GHz) rates." Phys. Rev. Lett. **102**, 113601 (2009).

32. H. Shin, J. A. Cox, A. Jarecki, A. Starbuck, Z. Wang, and P. T. Rakich, "Control of coherent information via on-chip photonic–phononic emitter–receivers," Nat. Comm. **6**, 6427 (2015).

33. T. Matsui, K. Nakajima, and F. Yamamoto, "Guided acoustic-wave Brillouin scattering characteristics of few-mode fiber," Appl. Opt. **54**, 6093-6097 (2015).

34. P. Dainese, P. St.J. Russell, G. S. Wiederhecker, N. Joly, H. L. Fragnito, V. Laude, and A. Khelif, "Raman-like light scattering from acoustic phonons in photonic crystal fiber," Opt. Express **14**, 4141-4150 (2006).

35. A. Butsch, M. S. Kang, T. G. Euser, J. R. Koehler, S. Rammler, R. Keding, and P. St. J. Russell, "Optomechanical nonlinearity in dual-nanoweb structure suspended inside capillary fiber," Phys. Rev. Lett. **109**, 183904 (2012).

36. P. K. A. Wai and C. R. Menyuk, "Polarization mode dispersion, decorrelation, and diffusion in optical fibers with randomly varying birefringence," J. Lightwave Technol. **14**, 148–157 (1996).

37. R. N. Thurston, "Elastic waves in rods and optical fibers," J. sound and vibration **159**, 441-467 (1992).

38. H. E. Engan, B. Y. Kim, J. N. Blake, and H. J. Shaw, "Propagation and optical interaction of guided acoustic waves in two-mode optical fibers." J. Lightwave Technol. **6**, 428-436 (1988).

39. B. A. Auld, Acoustic Fields and Waves in Solids, (Wiley and Sons, 1973).

40. H. Al-Raweshidy and S. Komaki, Radio over fiber technologies for mobile communications networks, (Artech House, 2002).

41. I. Gasulla and J. Capmany, "Microwave photonics applications of multicore fibers," IEEE Photonics Journal **4**, 877-888 (2012).

42. M. Wrage, P. Glas, D. Fischer, M. Leitner, D. V. Vysotsky, and A. P. Napartovich, "Phase locking in a multicore fiber laser by means of a Talbot resonator," Opt. Lett. **25**, 1436-1438 (2000).

43. E. J. Bochove, P. K. Cheo, and G. G. King, "Self-organization in a multicore fiber laser array," Opt. Lett. **28**, 1200-1202 (2003).

44. Y. Huo and P. K. Cheo, "Analysis of transverse mode competition and selection in multicore fiber lasers," J. Opt. Soc. Am. B **22**, 2345-2349 (2005).

45. L. Li, A. Schülzgen, S. Chen, V. L. Temyanko, J. V. Moloney, and N. Peyghambarian, "Phase locking and in-phase supermode selection in monolithic multicore fiber lasers," Opt. Lett. **31**, 2577-2579 (2006).

46. D. Levandovsky, M. Vasilyev, and P. Kumar, "Near-noiseless amplification of light by a phase-sensitive fibre amplifier," Pramana **56**, 281-285 (2001).

47. P. Frascella, S. Sygletos, F. C. Garcia Gunning, R. Weerasuriya, L. Gruner Nielsen, R. Phelan, J. O'Gorman and A. D. Ellis, "DPSK signal regeneration with a dual-pump nondegenerate phase-sensitive amplifier," IEEE Photon. Technol. Lett. **23**, 516-518 (2011).

48. J. P. Moore and M. D. Rogge, "Shape sensing using multi-core fiber optic cable and parametric curve solutions," Opt. Express **20**, 2967-2973 (2012).

49. Z. Zhao, M. A. Soto, M. Tang, and L. Thévenaz, "Distributed shape sensing using Brillouin scattering in multi-core fibers," Opt. Express **24**, 25211-25223 (2016).

50. X. S. Yao and L. Maleki, "Optoelectronic microwave oscillator," J. Opt. Soc. Am. B **13**, 1725-1735 (1996).

51. Y. London, H. H. Diamandi, and A. Zadok, "Electro-opto-mechanical radio-frequency oscillator driven by guided acoustic waves in standard single-mode fiber," submitted to Appl. Phys. Lett. Photon. (in review).